\documentclass[a4paper]{jpconf}
\usepackage{graphicx}
\usepackage{natbib}  
\usepackage{amsmath, gensymb}  
\usepackage{wrapfig}
\usepackage{dblfloatfix}
\usepackage{subcaption}
\usepackage{placeins}
\usepackage{color}
\usepackage{xcolor}
\usepackage{hyperref}
\usepackage{floatrow}
\usepackage{scrextend}

\begin{document}
\title{Expert Elicitation on Wind Farm Control}

\author{J. W. van Wingerden$^{\boldsymbol{\mathsf{1}}}$, P. A. Fleming$^{\boldsymbol{\mathsf{2}}}$, T. G\"{o}\c{c}men$^{\boldsymbol{\mathsf{3}}}$,
I. Eguinoa$^{\boldsymbol{\mathsf{4}}}$,
B.~M.~Doekemeijer$^{\boldsymbol{\mathsf{1}}}$,
K.~Dykes$^{\boldsymbol{\mathsf{3}}}$,
M. Lawson$^{\boldsymbol{\mathsf{2}}}$,
E. Simley$^{\boldsymbol{\mathsf{2}}}$,
J. King$^{\boldsymbol{\mathsf{2}}}$,
D. Astrain$^{\boldsymbol{\mathsf{4}}}$,
M. Iribas$^{\boldsymbol{\mathsf{4}}}$,
C. L. Bottasso$^{\boldsymbol{\mathsf{5}}}$,
J. Meyers$^{\boldsymbol{\mathsf{6}}}$,
S. Raach$^{\boldsymbol{\mathsf{7}}}$,
K. K\"{o}lle$^{\boldsymbol{\mathsf{8}}}$,
G.~Giebel$^{\boldsymbol{\mathsf{3}}}$}

\address{$^{\mathsf{1}}$Delft University of Technology, Delft, The Netherlands
$^{\mathsf{2}}$National Renewable Energy Laboratory, Golden, Colorado, USA
$^{\mathsf{3}}$Technical University of Denmark, Roskilde, Denmark
$^{\mathsf{4}}$CENER, Sarriguren, Spain
$^{\mathsf{5}}$Technical University Munich, Munich, Germany
$^{\mathsf{6}}$KU Leuven, Leuven, Belgium
$^{\mathsf{7}}$sowento, Stuttgart, Germany
$^{\mathsf{8}}$SINTEF Energy Research, Trondheim, Norway}

\ead{j.w.vanwingerden@tudelft.nl}

\begin{abstract}
Wind farm control is an active and growing field of research in which the control actions of individual turbines in a farm are coordinated, accounting for inter-turbine aerodynamic interaction, to improve the overall performance of the wind farm and to reduce costs. The primary objectives of wind farm control include increasing power production, reducing turbine loads, and providing electricity grid support services. Additional objectives include improving reliability or reducing external impacts to the environment and communities.
In 2019, a European research project (FarmConners) was started  with the main goal of providing an overview of the state-of-the-art in wind farm control, identifying consensus of research findings, data sets, and best practices, providing a summary of the main research challenges, and establishing a roadmap on how to address these challenges. Complementary to the FarmConners project, an IEA Wind Topical Expert Meeting (TEM) and two rounds of surveys among experts were performed. From these events we can clearly identify an interest in more public validation campaigns. Additionally, a deeper understanding of the mechanical loads and the uncertainties concerning the effectiveness of wind farm control are considered two major research gaps.
\end{abstract}

\section{Introduction}
Wind farm control (WFC) is an active and growing field of research in which the control actions of individual turbines in a farm are coordinated, accounting for inter-turbine aerodynamic interaction, to improve the overall performance of the wind farm and to reduce costs. The primary objectives of WFC include increasing power production, reducing turbine loads, and providing electricity grid support services. Additional objectives include improving reliability or reducing external impacts to the environment and communities.
In 2019, a European project (FarmConners\footnote{\scriptsize\label{note1}See \url{https://cordis.europa.eu/project/rcn/224282/factsheet/en} and \url{https://www.windfarmcontrol.info}.}) was started  with the main goal of providing an overview of the state of the art in WFC, identifying consensus of research findings, data sets, and best practices, providing a summary of the main research challenges, and establishing a roadmap on how to address these challenges. Complementary to the FarmConners project, an International Energy Agency (IEA) Wind Topical Expert Meeting (TEM) has similar objectives. In this paper these two initiatives are merged.

WFC can be categorized accordingly to three distinct technologies used to achieve the primary objectives identified above:
\begin{enumerate}
  \item The first technology is \textit{wake steering}, where wake interactions are modified by redirecting the wakes in the wind farm. This technique could be used to either increase power production by steering wakes away from downstream turbines or to reduce asymetric loading introduced by partial wakes. 
  \item The second technology is \textit{axial induction control}, where wake interactions and impacts are modified by derating upstream or uprating downstream turbines. Derating leads to a reduction in the structural loads of the derated and downstream turbines while uprating can increase power production. Both derating and uprating can provide the basis for supporting grid services such as active power control.
  \item The third technology is \textit{wake mixing}, where upstream turbines are dynamically uprated and downrated on short timescales to induce additional wake recovery, minimizing wake losses further downstream.
\end{enumerate}

The literature contains an abundance of WFC solutions that leverage one or more of these technologies for one or more of the aforementioned objectives \cite{Boersma2017b,KHEIRABADI201945,Knudsen2015}. A popular method of algorithm validation, much more cost-effective than field experiments, has been the use of dedicated wind tunnel experiments. The potential of wake steering has been demonstrated on multiple occasions. For example, Adaramola and Krogstad \cite{Adaramola2011}, Schottler et al. \cite{Schottler2016}, and Bartl et \textit{al.} \cite{Bartl2018} report gains of up to $12\%$ for wake steering in two-turbine arrays in wind tunnel experiments. Moreover, Campagnolo et \textit{al.} \cite{Campagnolo2016,Campagnolo2016b} and Park et \textit{al.} \cite{Park2016} report gains of up to $33\%$ for three-turbine arrays through wake steering in their wind tunnels. Bastankhah and Port\'e-Agel \cite{Bastankhah2019} report gains of up to $17\%$ for a five-turbine array using wake steering in a wind tunnel. Moreover, Campagnolo et \textit{al.} \cite{Campagnolo2016,Campagnolo2016b}, among others, tested the potential of axial induction control for power maximization in their wind tunnel reporting no net gains. The third concept, wake mixing, is a novel technology and thereby has only been validated to a limited degree in the literature. The concept demonstrated in simulation by Munters et \textit{al.} \cite{Munters:2018} was validated by Frederik et \textit{al.}~\cite{frederik2019} in a set of wind tunnel experiments. More recently, the \textit{helix} wake mixing concept was introduced by van Wingerden et \textit{al.}~\cite{ipcpatent}, where individual pitch control is used to trigger wake mixing~\cite{frederik2020}. This concept is still to be validated through scaled experiments.

A handful of publications have focused on validation of wind farm control algorithms through field experiments. Contrary to the findings of Campagnolo et \textit{al.} \cite{Campagnolo2016}, Van Der Hoek et \textit{al.} \cite{Hoek2019} demonstrate axial induction control in field trials on a commercial wind farm, showing a small but positive increase in the power production of the farm compared to baseline operation. Fleming et \textit{al.} \cite{Fleming2017} demonstrate wake steering on an offshore two-turbine array with positive results. Thereafter, Fleming et \textit{al.} \cite{Fleming2019,Fleming2020} demonstrate wake steering at an onshore two-turbine array surrounded by a complex topology. In these experiments, the benefits of wake steering are common in the data, but some losses are measured, too. Howland et \textit{al.} \cite{Howland2019} demonstrate wake steering at an onshore six-turbine array, showing significant gains in power production for particular situations (i.e., low wind speed, wake-loss-heavy wind directions, low turbulence levels), although they report that their net gain over annual operation appears insignificant compared to baseline operation. Doekemeijer et al. \cite{Doekemeijer2020} then demonstrate wake steering in a complex onshore wind farm, containing different turbine types and more complicated wake interaction. Their results largely agree with the other field experiments: both gains and losses in power production are measured. The authors conclude that, while the gains outnumber the losses, more research is necessary on the topic of wake steering. At large, it is clear that more research is necessary before WFC can be fully commercialized.

To accelerate the industrialization of WFC, the primal objectives of the European FarmConners project are to develop an overview of the:
\begin{enumerate}
  \item state of the industry: What are the current capabilities of commercial WFC solutions?
  \item state of the academia: What are the current research findings, including theoretical results, overview of mathematical models and their capabilities, and the findings of high-fidelity computer simulations, wind tunnel studies, and field trials?
  \item consensus of research findings and best practices: What topics have consensus and which questions remain unanswered in the research? Moreover, a secondary objective is to establish preferred nomenclature and define best practices for future research.
    \item research needs: Based on existing research and desired end goals, where is research most needed?
\end{enumerate}

In brief, for objective (i), the current commercial WFC solutions are dominated by steady-state wake steering where yaw misalignment is used to maximize the annual energy production (AEP). For objective (ii), recent articles \cite{Doekemeijer2019,Boersma2017b,KHEIRABADI201945} have reviewed the state of the research. Instead, this paper focuses on objectives (iii) and (iv) involving the consensus of research findings, best practices, and open research questions.

The paper is structured as follows. Section~\ref{sec:methodology} describes how information was gathered from leading academia and industry in the field of WFC. Section~\ref{sec:results} presents the findings. The paper is concluded in Section~\ref{sec:conclusion}.
\section{Methodology} \label{sec:methodology}
\begin{figure}[ht]
\centering
\includegraphics[width=.95\textwidth,trim=38mm 192mm 40mm 38mm,clip]{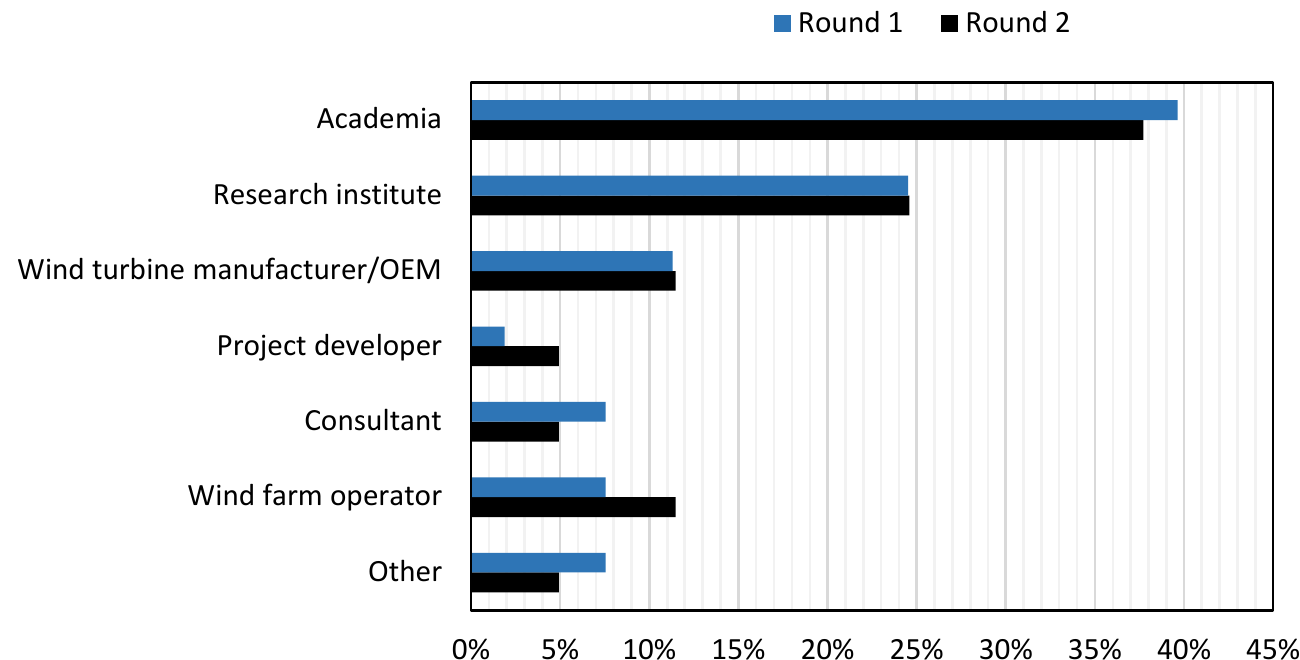}
\caption{Affiliation of the participants for the two rounds of WFC surveys}
\label{fig:Survey_participants}
\end{figure}

To meet the aforementioned objectives, IEA Wind TEM \#97 was held on the $25^{\textrm{th}}$ of September, 2019, in Amsterdam with 47 experts from academia and industry to discuss the current status of WFC. The TEM was preceded by two surveys about the current state of the art and research needs. The questionnaires were completed by over 50 WFC stakeholders. Figure~\ref{fig:Survey_participants} illustrates the affiliation of the participants for both iterations of the questionnaire. Although the input from academia and research institutes seems to have a higher share in the survey results, there is still considerable participation from the industrial and commercial stakeholders. 

There is a potential bias because of the high participation (more than 50\%) of academia and research institutes. Also, during the discussions in Amsterdam, it became apparent that the group, to a large degree, shared research interests, focusing mainly on flow modeling and aerodynamical turbine interaction. Experts working on control algorithms, the electrical components, and grid integration were underrepresented in the meeting. It can be assumed that the same group that attended the TEM also answered the survey, causing a certain bias toward, for example, power maximization by flow control. The potential effect of the participant background is discussed further in Section~\ref{sec:results}, where applicable. 

During the TEM, the aforementioned questionnaires, as well as the open outcomes of current research projects in the field, were used as a starting point for discussions during the meeting. Research on WFC has been separated into the aerodynamical topics and the aspects of grid integration. These discussions revealed that there was no uniform consensus among participants on the definition of ``grid services'' as mentioned in the questionnaire. While some attendees interpreted grid services as ``complying to grid codes,'' others interpreted them as a  ``provision of ancillary services'' and its related requirements. This should be considered when looking at the results in Section~\ref{sec:results}. Accordingly, establishing a clear definition is recommended in this paper, especially considering the expected evolution of grid codes and electrical market services as the penetration of wind energy in the energy mix increases over the coming years.


\section{Results} \label{sec:results}
The outcomes of the two-round survey are presented in this section. 

Before assessing the technical challenges of WFC, consensus must be reached on its definition. Therefore, the survey preceded with the question ``What is wind farm control?'' The results clearly point toward wake steering and axial induction control for power maximization and load mitigation (i.e., dealing with flow phenomena in the wind farm using mathematical models and handling uncertainty). Not surprisingly, because of the surveyed group, grid integration is rated as less important using this definition.

Moreover, one of the most pronounced outcomes of the survey is the common request for increased collaboration in WFC, mainly in terms of i) access to data, ii) availability of models, and iii) availability of field tests.  The main objectives of both the TEM in Amsterdam and the European FarmConners project address exactly that: more engaged collaboration through common dialogue platforms and open science principles. 

The survey is structured around three main queries, each elaborated next.

\subsection{\underline{What are the most important reasons for WFC?}}

\begin{description}
    \item[The most important benefits of WFC.]
    \begin{figure} [!ht]
        \centering
        \includegraphics[width=.95\textwidth]{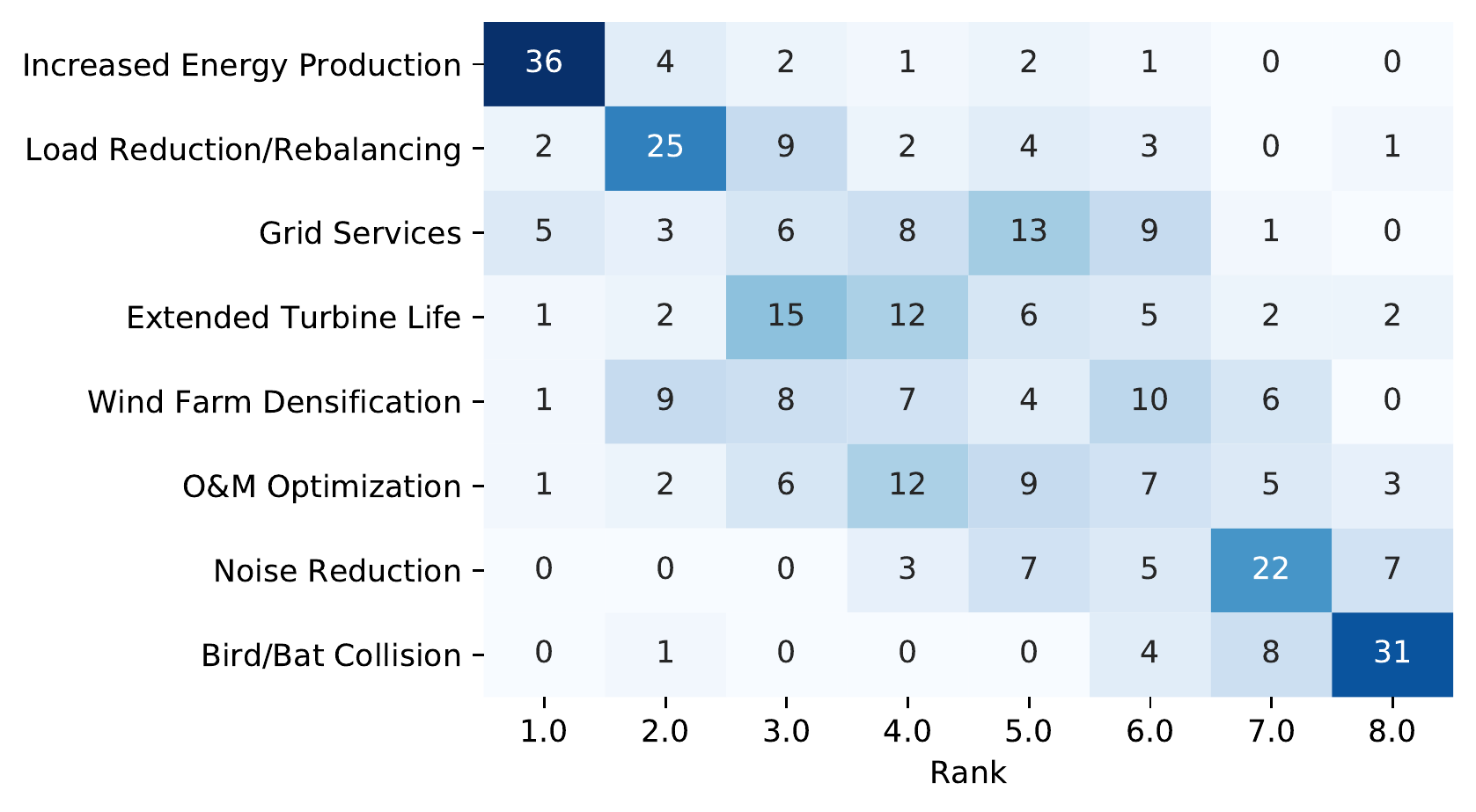}
        \caption{The most important benefits of WFC as perceived by the survey participants. Ranking on the x-axis refers to 1.0 -- most important, 8.0 -- least important}
        \label{fig:Benefits_WFC}
    \end{figure}
    The clear majority of the WFC community sees the increased energy production as the most important benefit of WFC implementation. Potential alleviation of turbine structural loads and lifetime extension follows as the second, where operations and maintenance (O\&M) optimization comes as third in importance. Figure~\ref{fig:Benefits_WFC} shows that mitigation of the environmental impacts such as noise reduction and bird/bat collision is ranked lowest in terms of the benefits that WFC can potentially provide. 
    
    \item[Expected AEP gain.] 
    Figure~\ref{fig:Expected_AEP_gain} shows the responses for the question ``How much of an increase in energy production is needed to justify implementation?'' In this figure, we see that around 60\% of the respondents believe that a gain of 1.0\% AEP gain is necessary to justify implementation. Looking at different stakeholder groups, however, we found that wind farm operators and turbine manufacturers had a lower threshold than 1.0\% while academia tended to have a higher threshold. As part of the maturity process in the field, this finding highlights the need to put the expected WFC benefits into better context by establishing more realistic but still commercially appealing predictions of potential improvements.
    \begin{figure}[b!]
        \centering
        \includegraphics[width=.7\textwidth,clip,trim=0 0 0 0mm]{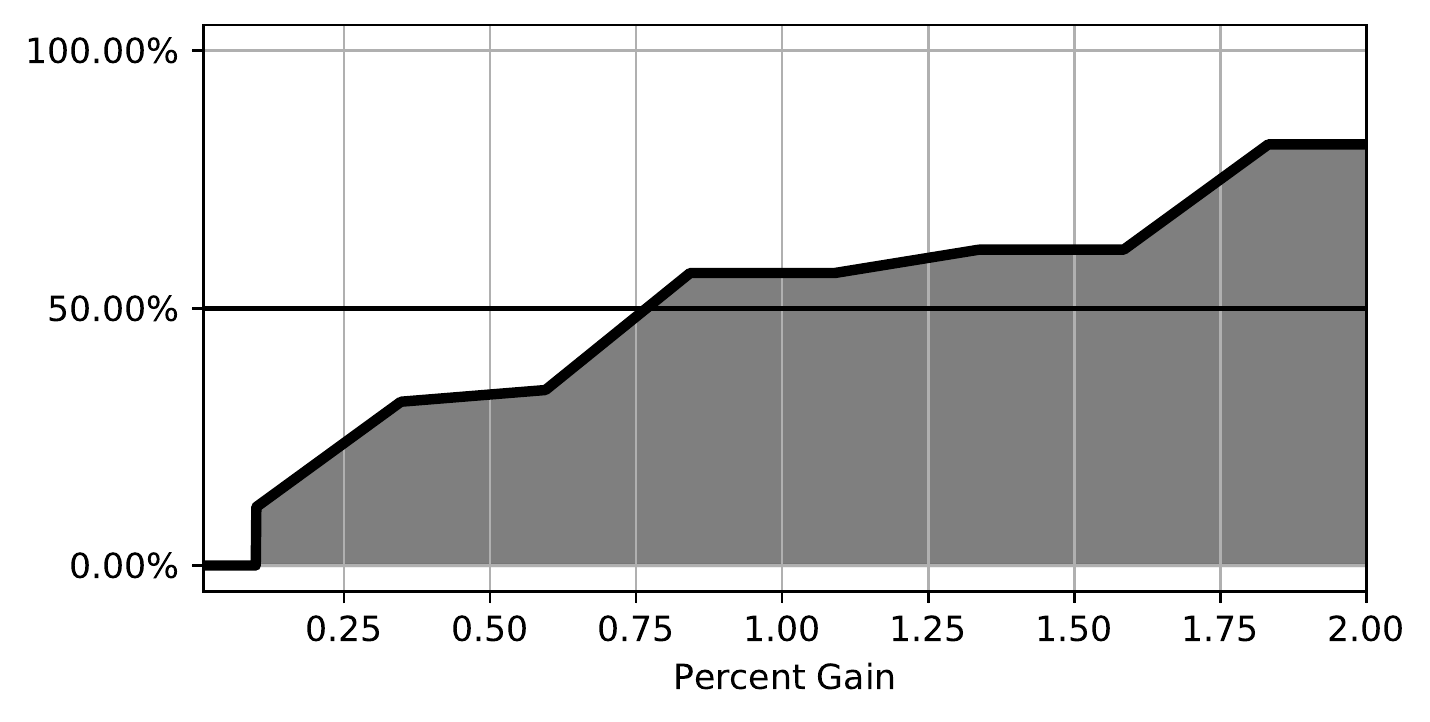}
        \caption{Distribution of responses to the question: How much of an increase in energy production is needed to justify implementation? The x-axis shows the percentage gain and the y-axis shows the cumulative percentage of questionnaire participants.}
        \label{fig:Expected_AEP_gain}
    \end{figure}
    
    \item[Plausibility of the current results in the literature.]
    \begin{figure}[b!]
        \begin{subfigure}{.49\textwidth}
            \centering
            \includegraphics[height=55mm]{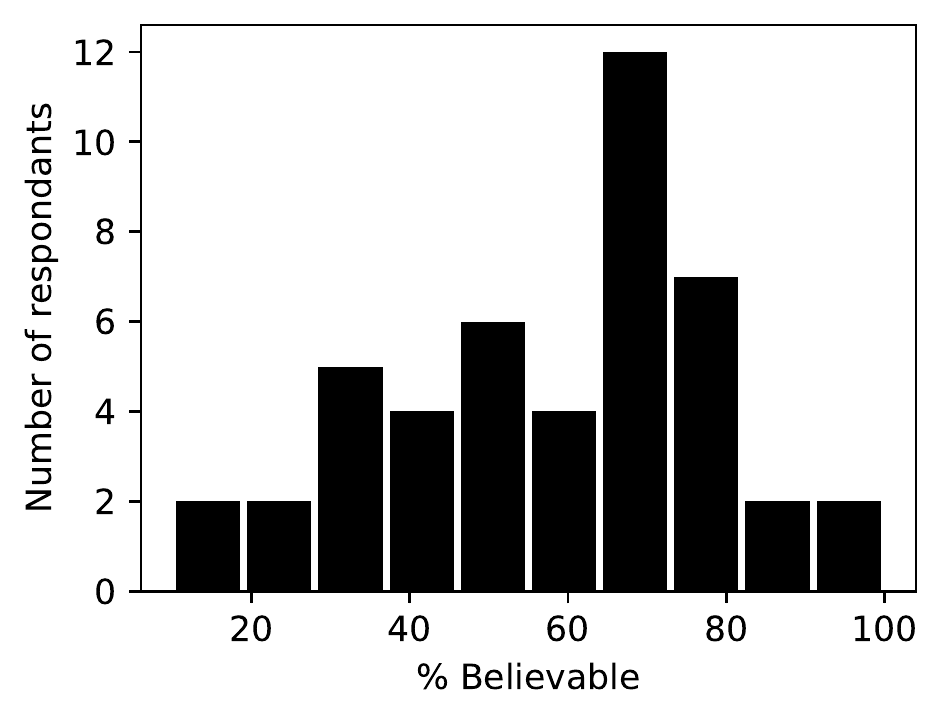}
        \end{subfigure}\begin{subfigure}{.49\textwidth}
            \centering
            \includegraphics[height=55mm]{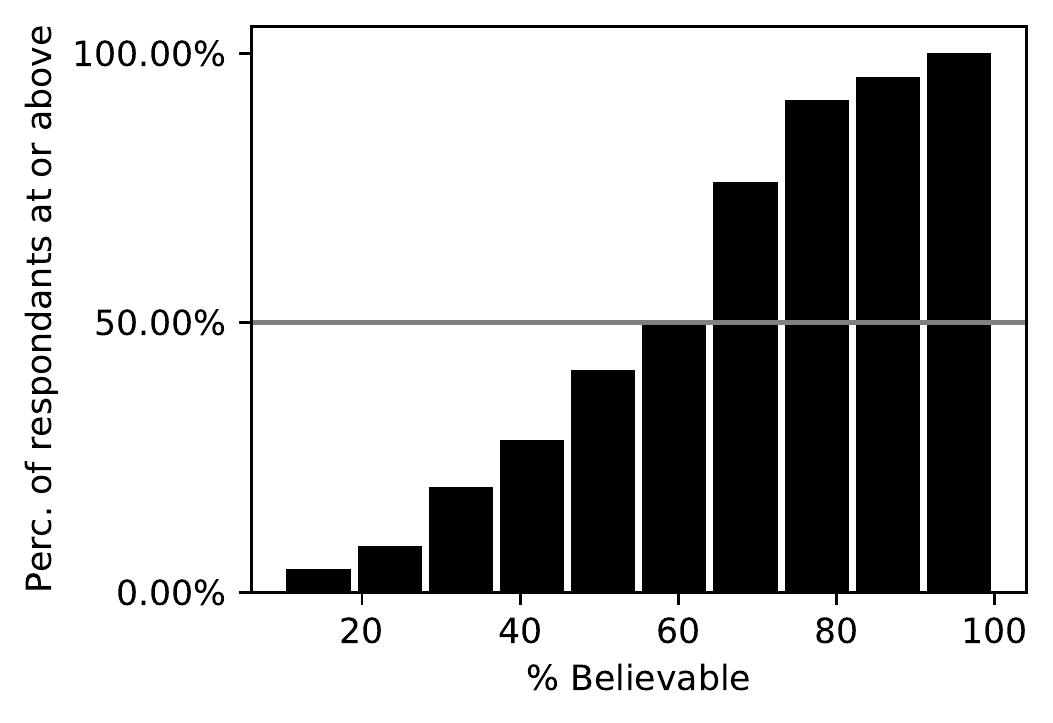}
        \end{subfigure}
        \caption{Distribution of responses to the question: On a scale of 0--100, how believable do you consider the reported energy yield gains of WFC methods? 100 -- highly believable. \textit{Left:} histogram of the responses; \textit{right:} cumulative distribution }
        \label{fig:plausibility_results}
    \end{figure}
    Figure~\ref{fig:plausibility_results} indicates the plausibility of the current benefits reported in the WFC literature by the stakeholders of the technology. It is seen that 50\% of the survey participants find the reported results to be more than 60\% believable, where the median of the answers is approximately 70\% plausibility.
\end{description}

\subsection{\underline{Where is consensus and where is disagreement?}}
\vspace{5pt}
Thereafter, the survey participants were asked to indicate their agreement to several statements in five categories from ``Strongly agree'' to ``Strongly disagree.'' These agreement classes are then compiled into percentage distributions, as listed in Table~\ref{tab:mostly_agree_table}. For example, 84\% of the participants agree that the WFC technology will be broadly adopted in the future, and 78\% of respondents highlight the need for standardization in the validation tools and methods.\\

\begin{table}[t!]
    \centering
    \def\arraystretch{1.3}
    \caption{Response to various statements in five categories from ``Strongly agree'' to ``Strongly disagree.'' This table shows the percentage distribution.}
    \label{tab:mostly_agree_table}    
    \noindent \begin{tabular}{p{115mm} r l} \hline \hline
     \textbf{Question} & \textbf{Agree} & \textbf{Disagree} \\
     WFC will be broadly adopted at some point in the future & 84\% & 0\% \\
     There is a lack of standard reference validation tools and methods to be openly used for certification and bankability purposes & 78\% & 8\% \\
     It is worth developing WFC if the only benefit is to AEP & 74\% & 5\% \\
     Wind farm flow control will be broadly adopted within 10 years & 66\% & 9\% \\
     There is a lack of reliable tools to evaluate the load impact of certain modes of operation in WFC (e.g., yaw misalignment) & 78\% & 15\% \\
     It is worth developing WFC if the only benefit is to reduce the turbine loads & 64\% & 11\% \\
     Atmospheric condition measurements (e.g., lidars or other equipment) should be used as real-time inputs in future WFC applications & 65\%  & 13\% \\
     WFC, as opposed to individual turbine control, is needed to provide grid services & 60\% & 8\% \\
     The theory of WFC is mature & 21\% & 50\% \\
     WFC is only applicable to newly designed wind farms & 3\% & 80\% \\ \hline \hline
    \end{tabular}
\end{table}

\begin{figure}[ht!]
        \centering
        \includegraphics[scale=.75]{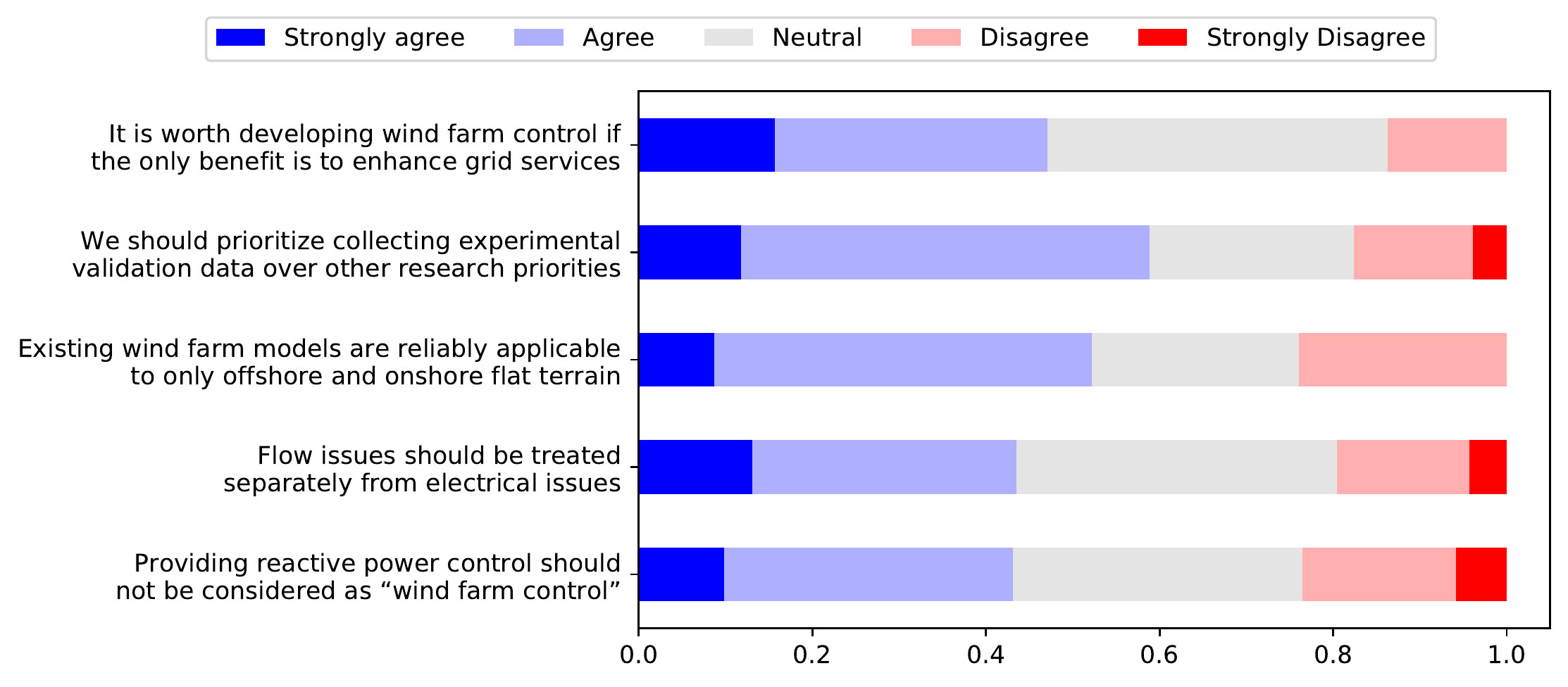}
        \caption{Statements to which more survey participants agree than disagree}
        \label{fig:MoreAgreeThanDisagree}
        \end{figure}

\begin{figure}[ht!]
        \centering
        \includegraphics[scale=.75]{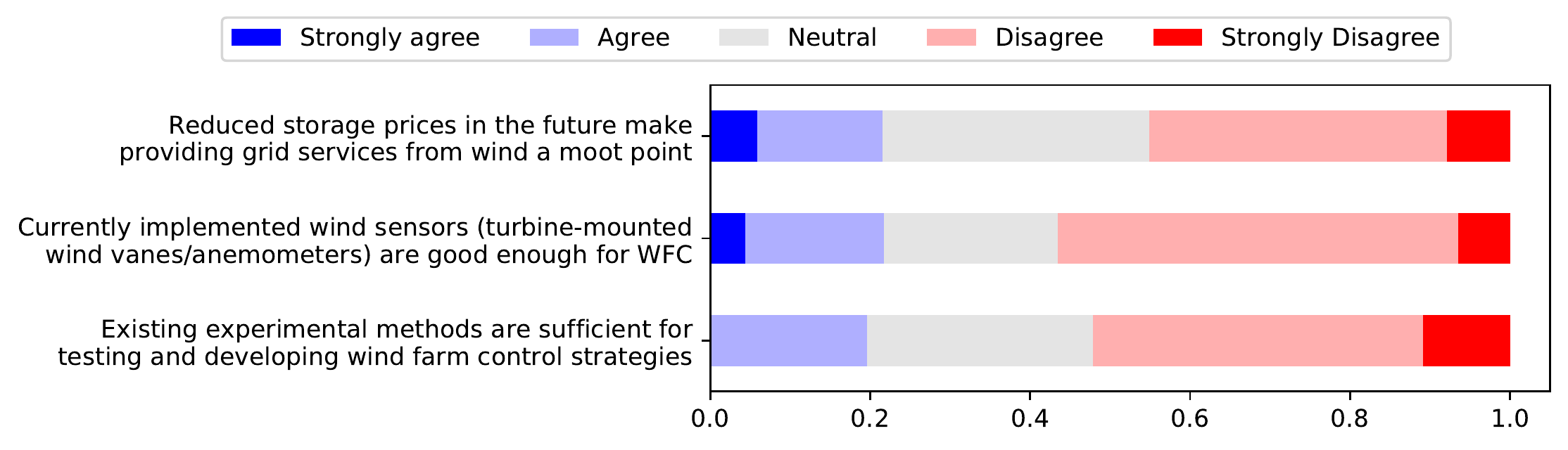}
        \caption{Statements to which more survey participants disagree than agree}
        \label{fig:MoreDisagreeThanAgree}
        \end{figure}

\begin{figure}[ht!]
        \centering
        \includegraphics[scale=.75]{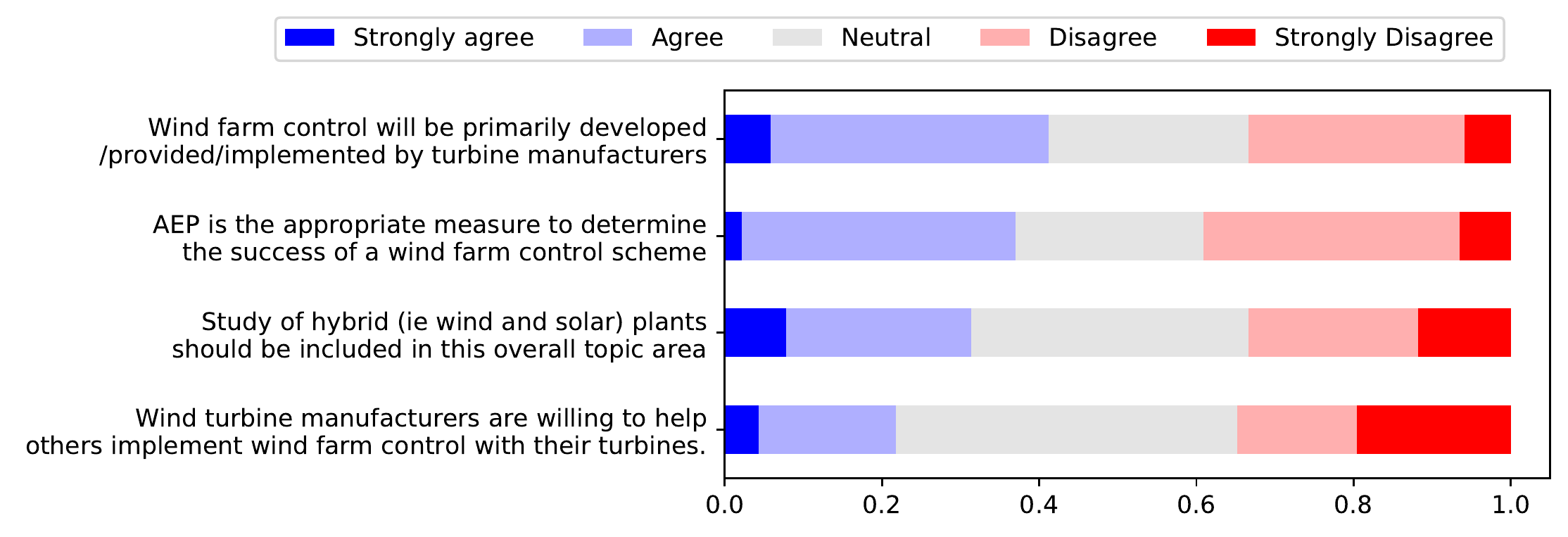}
        \caption{Statements without a clear trend among survey participants}
        \label{fig:Divided}
        \end{figure}

Figure~\ref{fig:MoreAgreeThanDisagree} shows statements to which most respondents agreed. Figure~\ref{fig:MoreDisagreeThanAgree} shows statements to which most respondents disagreed. Figure~\ref{fig:Divided} shows statements in which no consensus was found. It appears that WFC is deemed interesting as long as at least one of the objectives (power maximization, load mitigation, electricity grid services) can be achieved. Moreover, the respondents consider experimental validation a priority over other research topics, and current experimental methods are considered insufficient. It is interesting to note that there is no consensus on who will be primarily developing WFC algorithms---be it turbine manufacturers, academia, or an external entity.

\subsection{\underline{Prioritization of WFC research}} An important objective of the FarmConners project is defining implementation barriers and assigning research priorities. The corresponding research results are shown in Figures~\ref{fig:Barriers_implementation}~and~\ref{fig:Research_priorities}. 
     
\begin{description}
    \item[Validation.] The survey participants consider the lack of validation as one of the main barriers to WFC on an industrial scale, as shown in Figure~\ref{fig:Barriers_implementation}. The ranking of research priorities in Figure~\ref{fig:Research_priorities} shows a clear interest in public validation campaigns, which is in agreement with the findings from Figure~\ref{fig:MoreAgreeThanDisagree}, among others. Moreover, the survey shows that there is a strong need for consensus on widely acceptable validation methods. 
    \item[Mechanical loads.] Among the research barriers listed in Figure~\ref{fig:Barriers_implementation}, structural loads are consistently ranked to be of medium importance. Accordingly, a deeper understanding of mechanical loads is considered one of the important research gaps in Figure~\ref{fig:Research_priorities}.
    \begin{figure}[ht!]
        \centering
        \includegraphics[width=\textwidth,clip,trim=0 0 2cm 0]{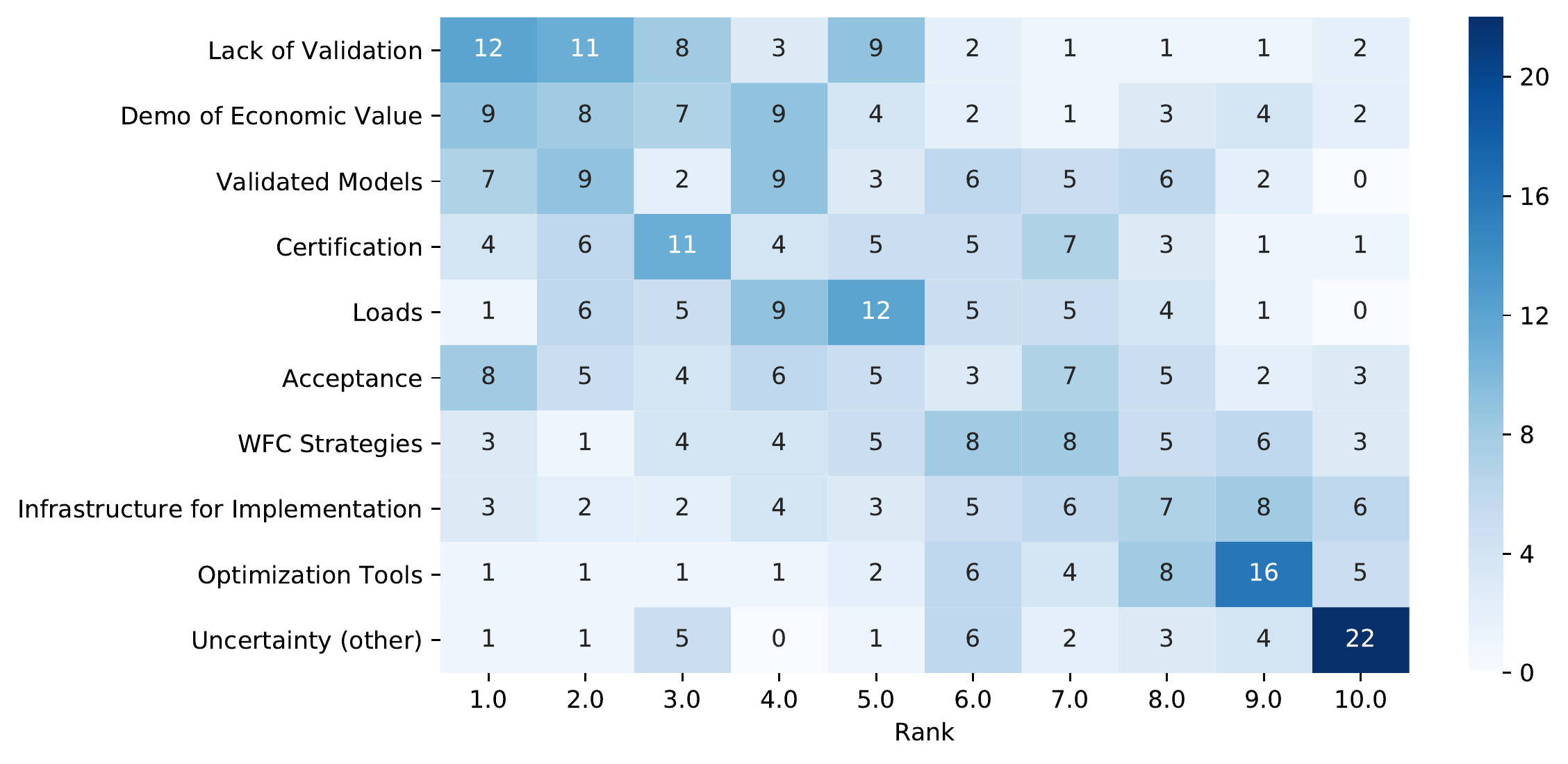}
        \caption{Barriers preventing implementation of WFC, ranked by the survey participants between 1 and 10 according to importance. 1 -- most important, 10 -- least important}
        \label{fig:Barriers_implementation}
        \end{figure}
    \item[Uncertainties concerning the effectiveness of WFC.] Finally, the understanding and quantification of statistical uncertainties is another major research gap according to the survey participants. This has only been explored to a very limited degree in WFC.
    \begin{figure}[ht!]
       \centering
        \includegraphics[width=\textwidth]{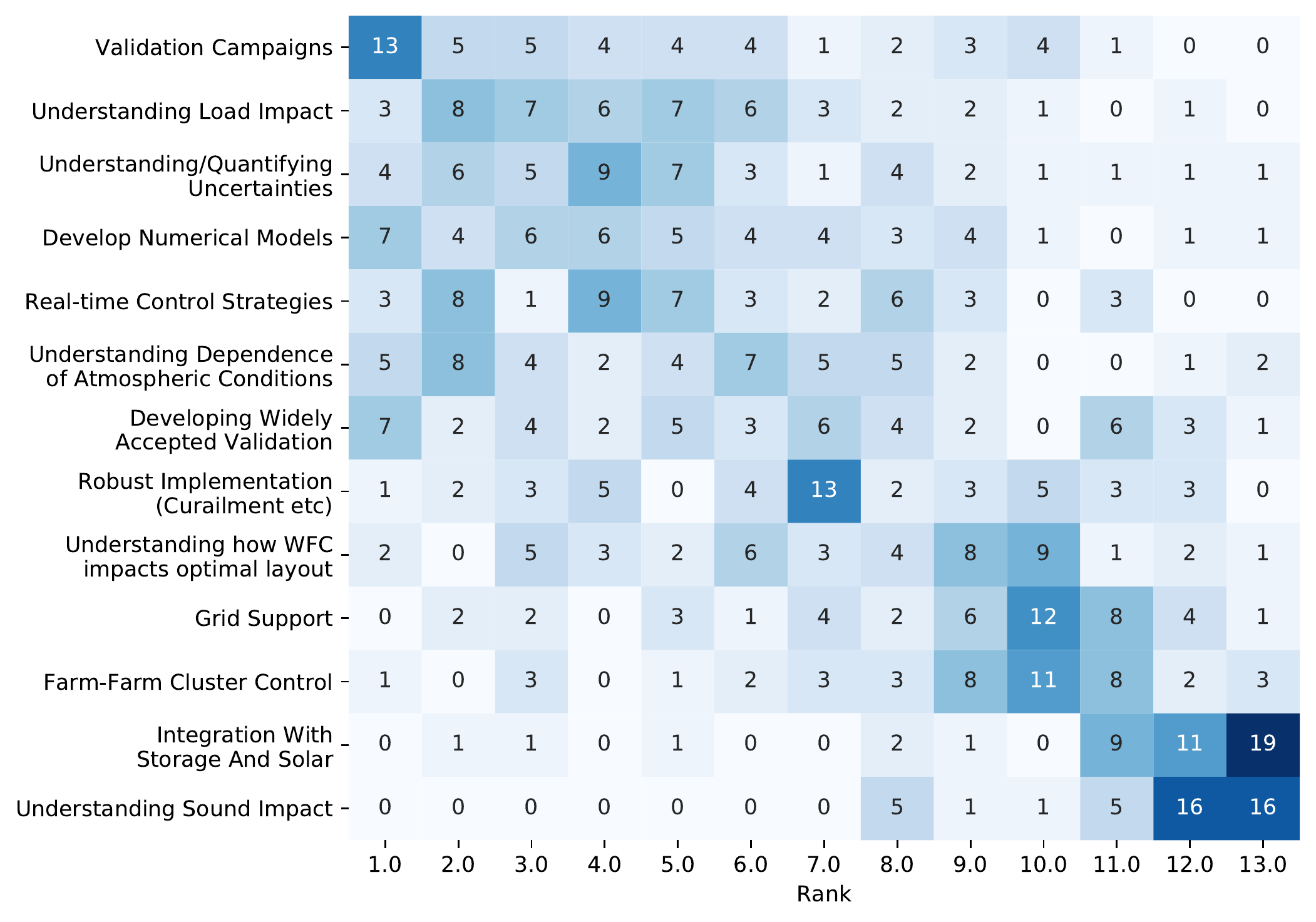}
        \caption{Research priorities for WFC according to the questionnaire; the different topics are ranked between 1 and 10 according to importance, 1 -- most important, 13 -- least important}
        \label{fig:Research_priorities}
    \end{figure}
\end{description}

The survey shows that the responding community considers topics such as the effect of WFC on farm layout optimization, control for farm-farm interaction, and integration with other renewables and aeroacoustics not to be current research priorities that will accelerate the implementation of industrial WFC.

\section{Conclusions}\label{sec:conclusion}
WFC is an active and growing field of research where consensus of research findings, best practices, and identification of open research questions are needed. This paper has shown current areas of consensus, areas with further disagreement, and research gaps that need to be addressed in the go-to-market path of WFC. Reaching the common standard definition of concepts, metrics, and tools is recommended according to the survey results and TEM discussions. Moreover, a stronger collaboration not only across institutions but also different disciplines working on the control of wind power plants is desirable and one of the declared goals of FarmConners. After all, this survey has provided interesting insight into the beliefs and assumptions driving WFC research. Taking the outcomes of the TEM into account, a follow-up survey focusing on industry will be conducted.


\section*{Acknowledgements}
\parbox{0.15\textwidth}{\includegraphics[height=3\baselineskip]
  {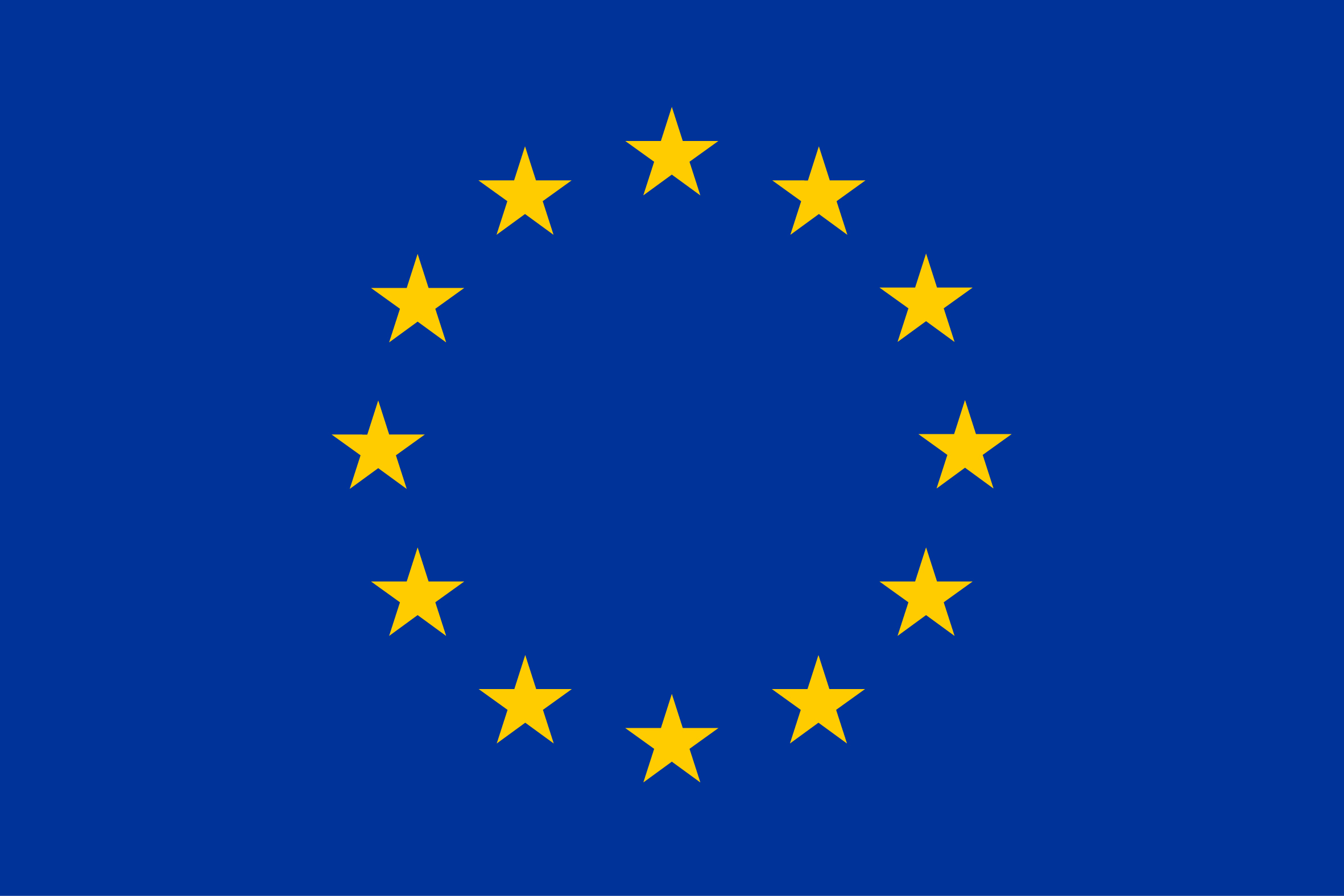}} \hfill
\parbox{0.85\textwidth}{A majority of the authors have received funding from the European Union’s Horizon 2020 Research and Innovation Programme under Grant Agreement No. 857844 as part of the FarmConners project. $^1$ has received funding by the Netherlands Organization for Scientiﬁc Research (NWO) as part of his VIDI grant with project number 17512}

\bibliographystyle{plain}
\bibliography{bibliography}
\end{document}